\documentclass[aps,pra,preprint,groupedaddress,floatfix]{revtex4-1}

\usepackage{graphicx}
\usepackage{hyperref}
\usepackage{amsmath,amsthm,amssymb}
\usepackage{booktabs}
\usepackage{floatrow}
\usepackage[caption=false]{subfig}

\begin{document}


\title{Stark resonance parameters for the $3a_{1}$ orbital of the
  water molecule}


\author{Susana Arias Laso}
\author{Marko Horbatsch}
\affiliation{Department of Physics and Astronomy, York University,
  Toronto, Ontario, Canada M3J 1P3}


\date{\today}

\begin{abstract}
  The Stark resonance parameters for the $3a_{1}$ molecular orbital of
  H$_{2}$O are computed by solving a system of partial differential
  equations in spherical polar coordinates. The starting point of the
  calculation is the quantum potential derived for this orbital from a
  single-center expanded Hartree-Fock orbital. The resonance positions
  and widths are obtained after applying an exterior complex scaling
  technique to describe the ionization regime for external fields
  applied along the two distinct $\hat{z}$ directions associated with
  the symmetry axis. The procedure thus avoids the computation of
  multi-center integrals, yet takes into account the geometric shape
  of a simplified molecular orbital in the field-free case.

\end{abstract}


\maketitle


\section{\label{sec:intro} Introduction}

Despite the complexity that the multi-center nature of the water
molecule entails, it has been the topic of numerous studies including
laser-induced ionization and high-harmonic
generation~\cite{PRL.107.083001, PRA.81.023412}, as well as electron
capture and ionization processes in ion-molecule
collisions~\cite{PRA.85.052713,PRA.86.022719,PRA.93.052705,
  PRA.93.032704,PRA.93.062706,PRA.87.032709,PRA.87.052710,Errea201517}. Most
calculations are within the framework of the independent electron
model and use a multi-center description of the
potential~\cite{PhysRevA.87.032709,Errea201517}. A strong motivation
to continue exploring this subject comes from the fundamental role
which ionization plays in radiation damage of biological tissue.

In a previous study of the H$_{2}$O valence orbitals exposed to strong
dc fields, we used an approach to determine the resonance parameters
for a given geometry of the orbitals without multi-center
integrals~\cite{PhysRevA.94.053413}. Based on the implementation of an
exterior complex scaling method, a system of partial differential
equations was solved numerically. The molecular potential was
expressed as a spherically symmetric effective potential obtained from
a single-center basis Hartree-Fock (HF)
calculation~\cite{MocciaJCP40III}. The ionization parameters for the
$1b_{1}$ and $1b_{2}$ molecular orbitals were explored over a range of
electric field strengths.

Here we extend the approach to study the dc Stark problem for the
$3a_{1}$ molecular orbital of H$_{2}$O. Given the orientation of this
orbital with respect to the plane in which the two protons are located
it is deemed necessary to go beyond the spherical effective potential
approximation which was used for the $1b_{1}$ and $1b_{2}$ orbitals.
This is accomplished by deriving a potential $V_{\rm{eff}}(r,\theta)$
for the $3a_{1}$ orbital from the single-electron Schr\"{o}dinger
equation with HF orbital wavefunction and energy supplied as known
quantities.

This paper is organized as follows: In
Sec.~\ref{sec:quantum_potential} the construction of the effective
potential $V_{\rm{eff}}(r,\theta)$ is presented. The required
asymptotic corrections applied to the electronic potential are given
in Sec.~\ref{sec:interpolation}, followed by a description of the
problem in terms of a system of partial differential equations in
Sec.~\ref{sec:pdeApproach}. Numerical results for the resonance
parameters are presented in Sec.~\ref{sec:results}, followed by
conclusions in Sec.~\ref{sec:conc}. Atomic units
($\hbar = m_{e} = e = 4\pi\epsilon_{0} = 1$) are used throughout.

\section{\label{sec:quantum_potential} Non-spherical effective
  potential derived from molecular orbitals}

The starting point for this work is the HF calculation of the H$_{2}$O
molecule in a single-center Slater orbital
basis~\cite{MocciaJCP40III}. Previously, we used the dominant parts of
the $1b_{1}$ and $1b_{2}$ orbitals, namely the $np_{x}$ and $np_{y}$
parts to derive spherically symmetric effective orbital-dependent
potentials and applied a Latter correction to guarantee the proper
asymptotic behavior for the respective
potential~\cite{PhysRev.99.510,PhysRevA.94.053413}.

Applying the same procedure to the $3a_{1}$ orbtital, i.e., retaining
the $np_{z}$ parts of the MO only leads again to a spherically
symmetric effective potential. Since we are interested in the response
of the orbital when applying an electric dc field along the symmetry
axis (i.e., the $z-$axis), there is an obvious deficiency: the two
protons (located in the $y-z$ plane) introduce a strong assymetry,
which leads to significant admixtures of $s-$type Slater orbitals in
the Moccia Slater-type orbitals (STO's)~\cite{MocciaJCP40III}.

\begin{figure}[t!]
  \centering
  \subfloat[Simplified $3a_{1}$ orbital\label{subfig:spherical}]
  {\includegraphics[width=0.25\textwidth]{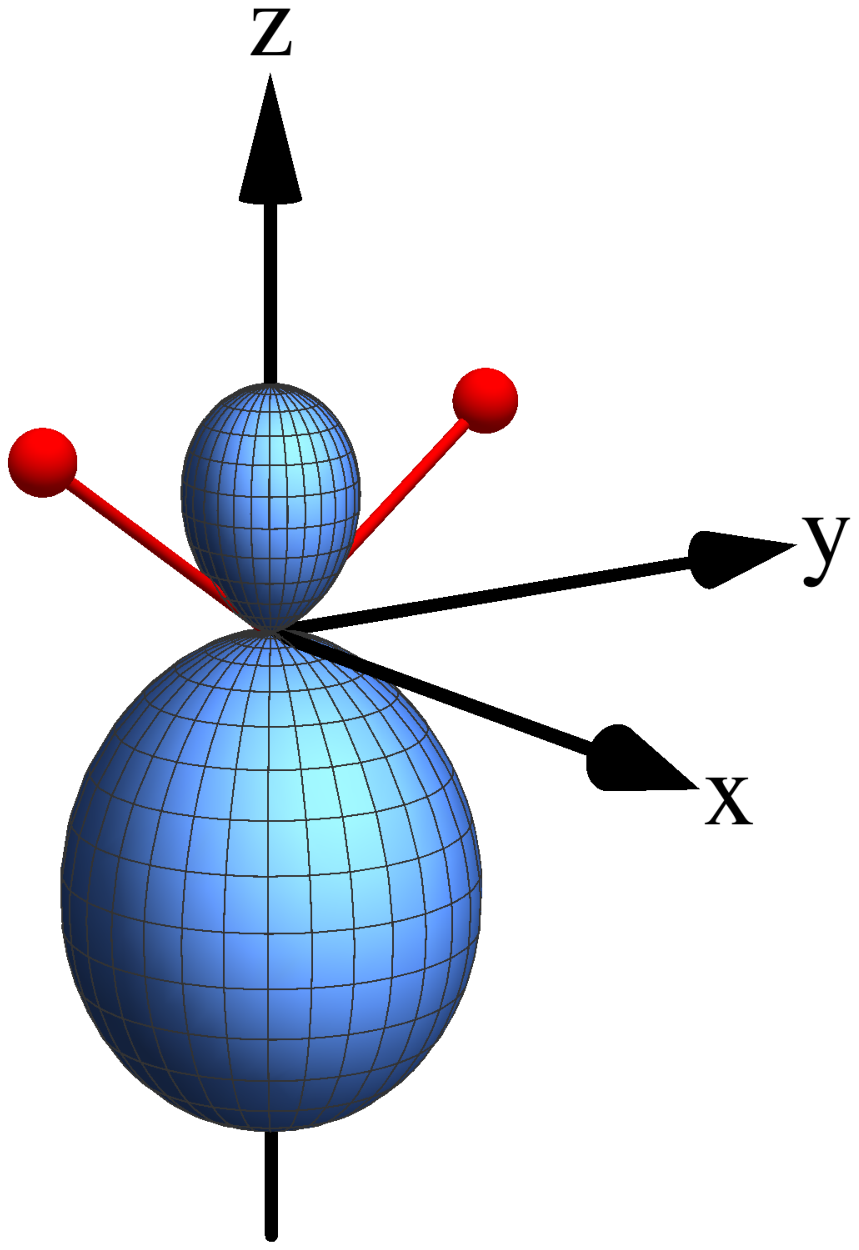}}\hspace{0.25\textwidth}
  \subfloat[Full Moccia $3a_{1}$ orbital\label{subfig:nonCentral}]
  {\includegraphics[width=0.25\textwidth]{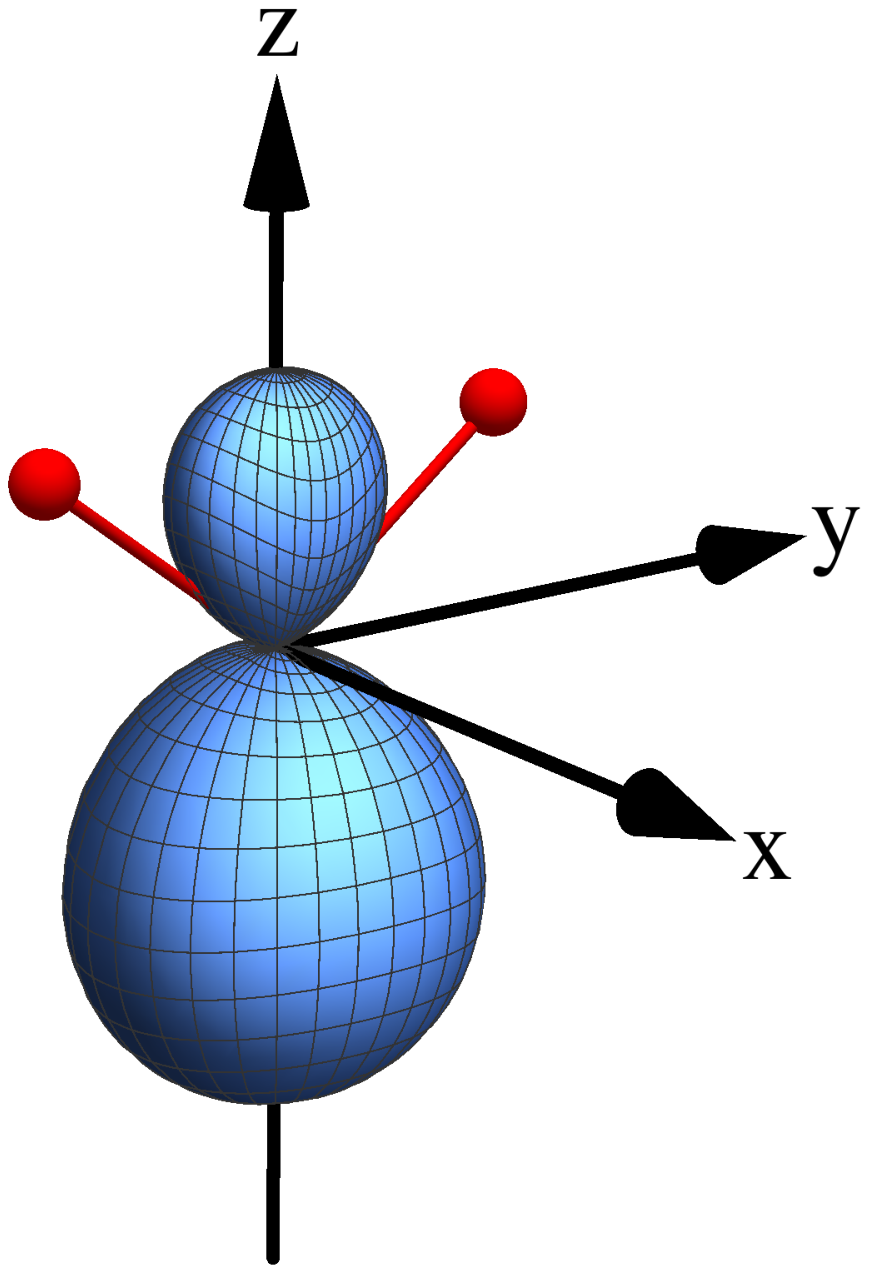}}
  \caption{Schematic display of the $3a_{1}$ molecular orbital (shown
    in blue along the $z$ axis) used to construct
    $V_{\rm{eff}}(r,\theta)$. The orbital obtained from a reduced
    expansion in STO's is shown in~(\ref{subfig:spherical}), and the
    complete Moccia orbital is shown in~(\ref{subfig:nonCentral}).
    Also indicated (in red in the $y-z$ plane) is the location of the
    protons. The $\hat{z}-$axis is the direction along which the
    external electric field of strength $F_{0}$ is applied.}
  \label{fig:3a1orbital}
\end{figure}

\begin{figure}[t!]
\includegraphics[width=0.49\textwidth]{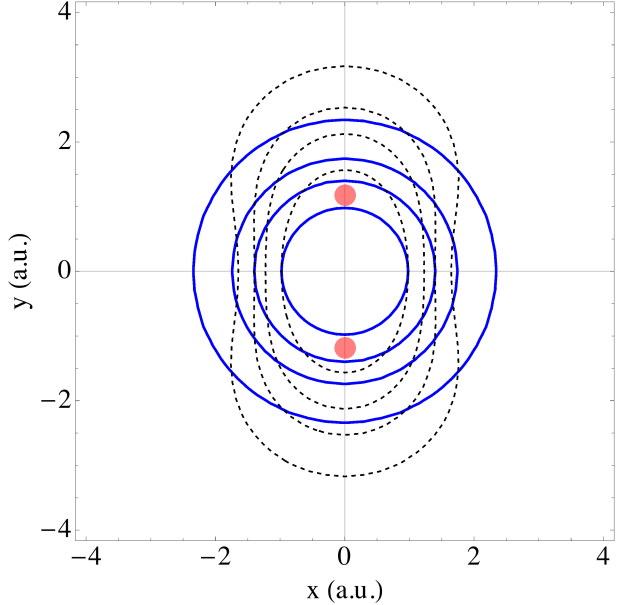}
\caption{Projections of the probability densities for the $3a_{1}$
  orbital on the $x-y$ plane. The simplified STO expansion is
  indicated as continuous blue lines, and the full Moccia expansion is
  indicated as black dashed lines. The protons are also indicated as
  red circles. The chosen contour values are $0.5, 0.3, 0.2, 0.1$
  starting from the innermost contour.}
\label{fig:contourxy}
\end{figure}

The proposed method to address this problem is to define a reduced
single-center Moccia wave function,
\begin{eqnarray}
\psi_{3a_{1}}(r,\theta) =
\sum_{n,l}{c_{nl0}\varphi_{nl0}(r,\theta)}.
\label{Moccia_expansion}
\end{eqnarray}
Here the $\varphi_{nl}(r,\theta)$ are Slater orbitals with $m=0$ for
the magnetic quantum number, and we limited the expansion to STO's of
$2s$ and $2p_{z}$ type. The parameters are given in
Table~\ref{tab:f_nlmMoccia} and three $2p_{z}$ orbitals are mixed with
three $2s-$type orbitals. This set of coefficients represents a
reduced selection of the expansion parameters given by Moccia for the
ground state of the water molecule~\cite{MocciaJCP40III} also shown in
Table~\ref{tab:f_nlmMoccia}.

The probability densities for the $3a_{1}$ orbital as obtained from
the reduced expansion~(\ref{Moccia_expansion}) and from the Moccia
self-consistent results are shown in Figures~\ref{subfig:spherical}
and~\ref{subfig:nonCentral} respectively. The protons (in red) defined
in the $y-z$ plane. As Fig.~\ref{subfig:spherical} indicates, the
contributions to the density of the $2s-$type states reproduce the
proper dependence of the $3a_{1}$ probability density with the polar
angle $\theta$, as the broader hump is located on the negative $z$
axis in the same way that the complete Moccia representation
illustrates in Fig.~\ref{subfig:nonCentral}.

In order to illustrate the fraction of the full Moccia expansion that
our reduced wave function~(\ref{Moccia_expansion}) represents, the
projections of the probability densities over the $x-y$ plane are
shown as contours of constant density in Figure~\ref{fig:contourxy},
for the height where the protons are located. From the complete Moccia
representation of the $3a_{1}$ MO (in dashed lines), one observes that
the location of the protons (shown as red circles) has an influence on
the shape of the upper lobe in the probability density, i.e., it
introduces dependence on the azimuthal angle $\varphi$. In our
simplified expansion, where only $l=0,1$ and $m=0$ symmetrical parts
were included (shown with solid lines), the probability density misses
to represent the proper azimuthal dependence that follows from the
$m\neq 0$ parts.

The non-spherical effective potential corresponding to the STO
expansion~(\ref{Moccia_expansion}), $V_{\rm{eff}}(r,\theta)$, is
obtained from the Schr\"{o}dinger equation in spherical polar
coordinates,
\begin{eqnarray}
  \left[-\frac{1}{2}\nabla^{2} +
  V_{\rm{eff}}(r,\theta)\right]\psi_{3a_{1}}(r,\theta) =
  E_{3a_{1}}\psi_{3a_{1}}(r,\theta).
\label{sch_eq3d}
\end{eqnarray}
For given $E_{3a_{1}}$ and $\psi_{3a_{1}}(r,\theta)$ it is
straightforward to solve~(\ref{sch_eq3d}) for
$V_{\rm{eff}}(r,\theta)$. In order to use this potential to define a
Hamiltonian for the $3a_{1}$ orbital in an electric field an
asymptotic Latter correction needs to be applied.

\begin{table}[t]
\centering
\caption{\label{tab:f_nlmMoccia} Expansion coefficients and non-linear
  coefficients for the $3a_{1}$ MO. The parameters used in our
  reduced STO expansion are indicated as included.}
\begin{ruledtabular}
\begin{tabular}{lrrr}
\toprule
$(n,l,m)$ & & $c_{nlm}$ & $\zeta_{i}$ \\
\midrule[0.25pt]
 $(1,0,0)$ & excluded & $-0.00848$ & $12.600$ \\
 $(1,0,0)$ & excluded & $0.08241$ & $7.450$ \\
\midrule[0.25pt]
 $(2,1,0)$ & included & $0.79979$ & $1.510$  \\
 $(2,1,0)$ & included & $0.00483$ & $2.440$  \\
 $(2,1,0)$ & included & $0.24413$ & $3.920$  \\
 $(2,0,0)$ & included & $-0.30752$ & $2.200$ \\
 $(2,0,0)$ & included & $-0.04132$ & $3.240$ \\
 $(2,0,0)$ & included & $0.14954$ & $1.280$ \\
\midrule[0.25pt]
 $(3,2,0)$ & excluded & $0.05935$ & $1.600$ \\
 $(3,2,0)$ & excluded & $0.00396$ & $2.400$ \\
 $(3,2,2)$ & excluded & $-0.09293$ & $1.600$ \\
 $(3,2,2)$ & excluded & $0.01706$ & $2.400$ \\
 $(4,3,0)$ & excluded & $-0.01929$ & $1.950$ \\
 $(4,3,2)$ & excluded & $-0.06593$ & $1.950$ \\
\bottomrule
\end{tabular}
\end{ruledtabular}
\end{table}

\subsection{\label{sec:interpolation} Interpolation and Latter correction of the
  effective potential}

The non-central effective potential, $V_{\rm{eff}}(r,\theta)$, leads
no longer to an orbital of $(l,m)$ symmetry, i.e., $2p_{z}$. This
reflects the geometry of the problem as a consequence of the location
of the protons. The use of this more general potential implies that
the Latter criterium~\cite{PhysRev.99.510}, which ensures the proper
asymptotic behavior of the potential, is not as straightforward to
implement as in the case of the spherical potential were the
correction applies beyond a determined $r$
value~\cite{PhysRevA.94.053413}. Now the correction must be
implemented in the $r-\theta$ plane, by defining a $\theta-$dependent
boundary beyond which the potential obtained from~(\ref{sch_eq3d})
rises above $-1/r$ in the asymptotic region.

We fix the $\theta$ coordinate at two extreme positions, such as
$\theta = 0$ and $\pi$, to find the corresponding $r$ values, $r_{0}$
and $r_{\pi}$, for which $V_{\rm{eff}}(r,\theta) = -1/r$ is satisfied,
and then interpolate between them by introducing a
$\theta-$dependent function. We use the function
\begin{eqnarray}
r_{\rm{match}}(\theta) & = & \bar{r} - (r_{\pi} - \bar{r})\cos\theta,
\label{eq:rMatch}
\end{eqnarray}
where $\bar{r}=(r_{0}+r_{\pi})/2$. With this approach we redefine the
effective potential to be the non-central potential derived from the
reduced Moccia wave function using Eq.~(\ref{sch_eq3d}) when
$r < r_{\rm{match}}(\theta)$, and $-1/r$ otherwise.

The weighted functions used to construct the Moccia
orbitals~\cite{MocciaJCP40III} imply a potential difficulty in our
problem. Since these functions are not exact solutions of the
Schr\"{o}dinger equation but were obtained from the variational
principle by implementing a self-consistent calculation
~\cite{MocciaJCP40I}, there may be regions in the $(r,\theta)$ domain
where $\psi_{3a_{1}}(r,\theta)$ vanishes, whereas its second
derivative remains finite; this produces a nodal line in the
electronic potential. Thus finding a potential for which our
approximate wave function satisfies a Schr\"{o}dinger equation
represents an intricate problem.

It turns out that the nodal region is so narrow that when solving the
Schr\"{o}dinger equation the kinetic energy term dominates and it is
possible to obtain a solution that remains close to that obtained by
the Hartree-Fock method~\cite{MocciaJCP40III}, regardless of the fact
that there is a region where the effective potential might diverge.

The probability density exhibits two humps indicating the positions of
the protons, which is consistent with Figure~\ref{fig:3a1orbital}, and
the effects of the mixing with the $s-$state.

\begin{figure}[t!]
\sidesubfloat[]
{\includegraphics[width=0.225\textwidth]{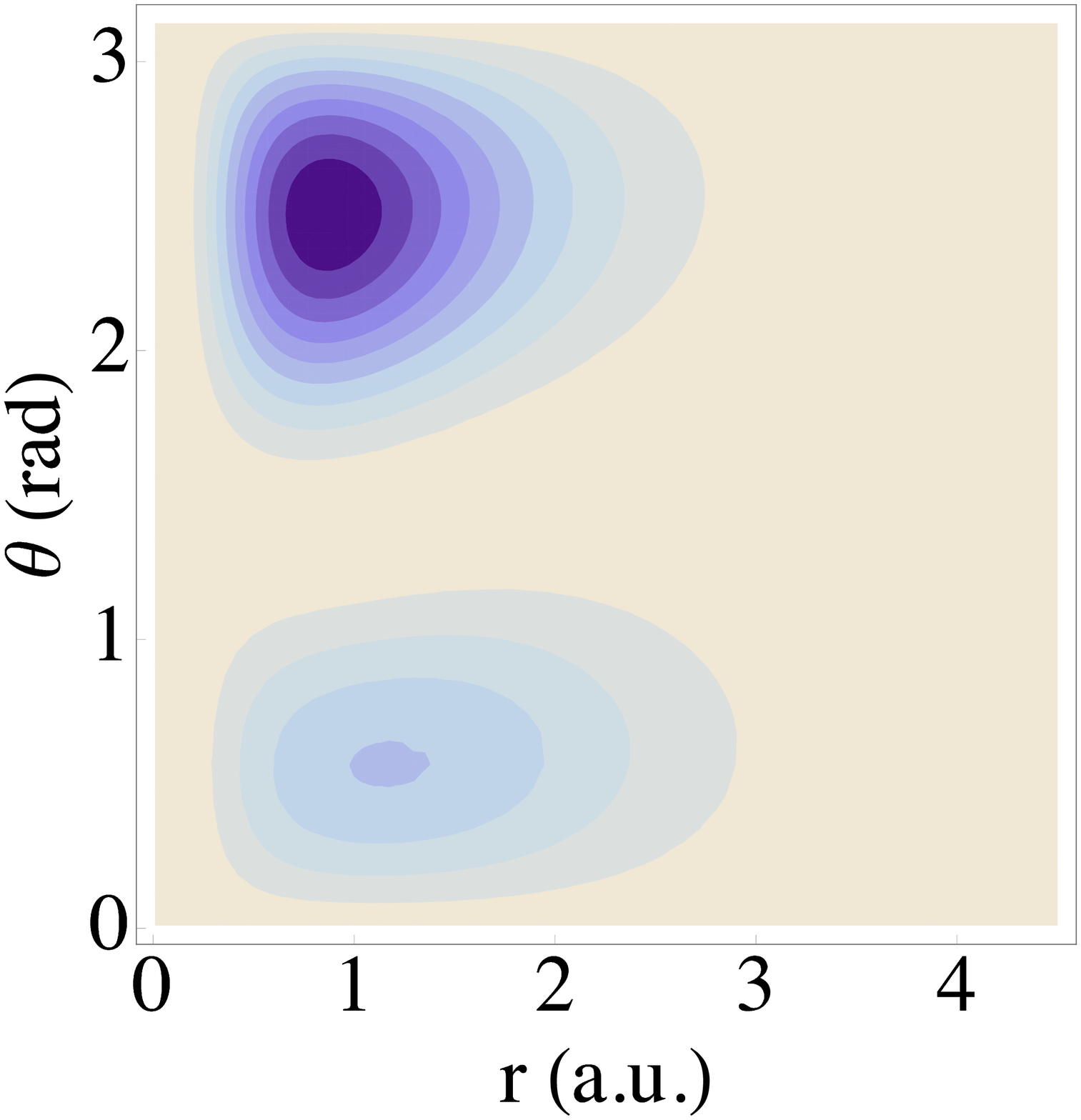}\label{Moccia32s}}
\sidesubfloat[]
{\includegraphics[width=0.253\textwidth]{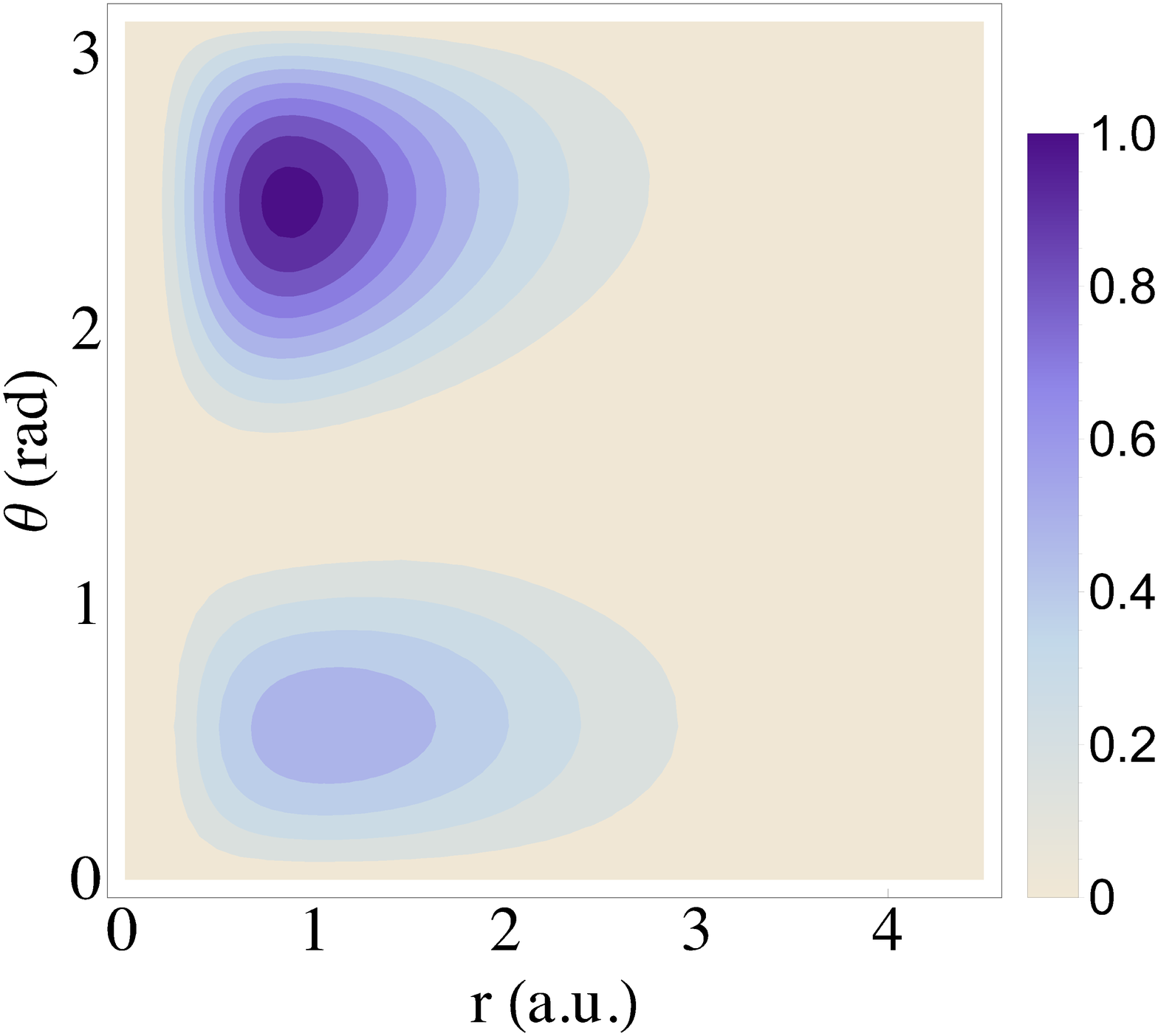}\label{intp32s}}
\caption{Contour plots of the probability density for the $3a_{1}$
  molecular orbital. The orbital density constructed from the reduced
  STO expansion is shown in~(\ref{Moccia32s}), while the solution
  obtained from the non-spherical $V_{\rm{eff}}(r,\theta)$ with Latter
  correction is shown in~(\ref{intp32s}).}
\label{fig:density_contours}
\end{figure}

One may argue that one of the reasons this nodal region in the
potential does not have a negative impact on the results is due to the
way the $3a_{1}$ orbital responds to the effective potential by
avoiding this region, its probability density being distributed as
shown in Figure~\ref{fig:density_contours}. We implement a numerical
interpolation of $V_{\rm{eff}}(r,\theta)$ in order to ensure it
continues smoothly over this problematic region.

The interpolation is achieved by collecting data from the evaluation
of the potential on two sections of the $(r,\theta)$ grid in the
vicinity of the nodal line, where the potential evaluates to finite
values. Then a numerical interpolation was carried out between those
regions in order to obtain a continuous function,
$V_{\rm{eff}}^{\rm{intp}}(r,\theta)$, on the two-dimensional grid. The
Latter correction is applied to the interpolated potential and the
effective potential is defined according to~(\ref{eq:rMatch}):
\begin{eqnarray}
  V_{\rm{eff}}(r,\theta) & = & \left.
  \begin{cases}
    V_{\rm{eff}}^{\rm{intp}}(r,\theta) & \text{for } r < r_{\rm{match}}(\theta) \\
    -1/r & \text{for } r > r_{\rm{match}}(\theta)
  \end{cases}
\right\}.
\label{eq:latterVeff}
\end{eqnarray}

Figure~\ref{Moccia32s} shows the probability density for the $3a_{1}$
MO as a contour plot in the $r-\theta$ plane as obtained from the
reduced Moccia expansion in Slater-type
orbitals~(\ref{Moccia_expansion}). Fig.~\ref{intp32s} shows the same
for the solution of the Schr\"{o}dinger equation~(\ref{sch_eq3d})
using the interpolated $V_{\rm{eff}}(r,\theta)$, given in
Eq.~(\ref{eq:latterVeff}), with the Latter
correction~\cite{PhysRev.99.510} applied in the asymptotic $r-$region.

\floatsetup[figure]{style=plain,subcapbesideposition=top}

The effective potential~(\ref{eq:latterVeff}) results in the
probability density shown in Fig.~\ref{intp32s} and yields an orbital
energy of $-0.5579\ \rm{a.u.}$ for the $3a_{1}$ MO, with a relative
change of $0.32\%$ in comparison with the self-consistent result of
Moccia~\cite{MocciaJCP40III} of $-0.5561\ \rm{a.u.}$

As Fig.~\ref{intp32s} indicates, the implementation of the Latter
correction to the orbital-dependent potential obtained from
Eq.~(\ref{sch_eq3d}), introduces a slight re-adjustment of the
density, with a somewhat higher probability density in the region
$0<\theta<\pi/2$. Since the Latter correction imposes an upper bound
of $-1/r$ in the effective potential beyond some $\theta-$dependent
boundary, this transformation in the effective potential establishes a
softer tail for the orbital, which gives rise to the probability
density re-distribution observed in Fig.~\ref{intp32s} vs
Fig.~\ref{Moccia32s}.

\subsection{\label{sec:pdeApproach} PDE in spherical polar coordinates}

The problem of describing the ionization regime of the $3a_{1}$ MO
under an external dc field applied along the orientation axis of the
orbital is expressed in terms of a system of partial differential
equations in spherical polar coordinates~\cite{PhysRevA.94.053413}. A
non-hermitian Hamiltonian is obtained as a result of applying exterior
complex scaling~\cite{PhysRevA.94.053413, complexScaling,
  complexScalingBaslev, complexScalingSimon,ecsSimon} to the radial
coordinate, where the $r-$coordinate is extended into the complex
plane by the phase function $\chi(r)$, $r\to r\exp[i\chi(r)]$. The
phase function $\chi(r)$ evolves smoothly from small values at $r=0$
to $\chi_{s}$ at large values of $r$ in the asymptotic region of the
effective potential where the potential is spherically symmetric and
purely Coulombic. The gradual increment of the scaling function is
implemented by the same function as used in~\cite{PhysRevA.94.053413}
as
\begin{eqnarray}
\chi(r) & = & \frac{\chi_{\rm{s}}}{1+\exp[-\frac{1}{\Delta r}(r-r_{\rm{s}})]},
\label{ecs_theta}
\end{eqnarray}
where the parameters $r_{\rm{s}}$ and $\Delta r$ were chosen for the
function $\chi(r)$ to rise smoothly from nearly zero to
$\chi_{\rm{s}}$ at $r-$values just outside where the Latter correction
is applied, i.e., $r_{\rm{s}}>r_{\rm{match}}$.

Exterior complex scaling again leads to a system of coupled partial
differential equations~(\ref{pde_system}), where the $R(I)$ labels
indicate the real and imaginary parts respectively due to the
coordinate mapping into the complex plane.

\begin{widetext}
\begin{eqnarray}
  -\frac{1}{2}\frac{\partial^{2}\psi_{R}}{\partial r^2}-\frac{1}{2r^2}
  (\frac{\cos\theta}{\sin\theta}\frac{\partial\psi_{R}}{\partial\theta}+
  \frac{\partial^{2}\psi_{R}}{\partial\theta^2}) \nonumber\\
  +[\frac{m^2}{2r^2\sin^2\theta}+V_{\rm{eff}}^{R}(r,\theta)c_{2}
  -V_{\rm{eff}}^{I}(r,\theta)s_{2}
  -E_{R}c_{2}+E_{I}s_{2}+
  F_{0}r\cos\theta c_{3}]\psi_{R} \nonumber\\
  +[-V_{\rm{eff}}^{R}(r,\theta)s_{2}-V_{\rm{eff}}^{I}(r,\theta)c_{2}
  +E_{R}s_{2}+
  E_{I}c_{2}-F_{0}r\cos\theta
  s_{3}]\psi_{I} & = & 0, \nonumber\\
  \vspace{1cm}
  -\frac{1}{2}\frac{\partial^{2}\psi_{I}}{\partial r^2}-\frac{1}{2r^2}
  (\frac{\cos\theta}{\sin\theta}\frac{\partial\psi_{I}}{\partial\theta}+
  \frac{\partial^{2}\psi_{I}}{\partial\theta^2}) \nonumber\\
  +[\frac{m^2}{2r^2\sin^2\theta}+V_{\rm{eff}}^{R}(r,\theta)c_{2}
  -V_{\rm{eff}}^{I}(r,\theta)s_{2}
  -E_{R}c_{2}+E_{I}s_{2}+
  F_{0}r\cos\theta c_{3}]\psi_{I} \nonumber\\
  +[V_{\rm{eff}}^{R}(r,\theta)s_{2}+V_{\rm{eff}}^{I}(r,\theta)c_{2}
  -E_{R}s_{2}-E_{I}c_{2}+
  F_{0}r\cos\theta
  s_{3}]\psi_{R} & = & 0.
\label{pde_system}
\end{eqnarray}
\end{widetext}

The system of equations~(\ref{pde_system}) was solved numerically on a
two-dimensional grid defined in $(r,\theta)$ coordinates. The domains
of $r$ and $\theta$ values were restricted to the intervals
$r\in[\epsilon, r_{\rm{max}}]$ and
$\theta\in[\eta,\theta_{\rm{max}}]$, with typical values
$\epsilon = \eta = 10^{-2}\ \rm{a.u.}$,
$r_{\rm{max}} = 28\ \rm{a.u.}$, and $\theta_{\rm{max}} = \pi-\eta$. In
the limit of low field strengths, i.e., $F_{0} = 0.05, 0.06$, the
value of $r_{\rm{max}}$ was increased to $40\ \rm{a.u.}$ in order to
ensure the outer turning points lie inside the grid, as the tunneling
barrier extends to larger $r$.

The problem of finding a solution of the Schr\"{o}dinger equation for
the $3a_{1}$ molecular orbital with contributions of $2s$ and
$2p-$type states requires a set of boundary conditions that describes
the properties of the orbital on the grid. In contrast with the
$m=\pm\ 1$ solutions obtained for the $1b_{1}$ and $1b_{2}$ MO's of
H$_{2}$O~\cite{PhysRevA.94.053413}, Neumann boundary conditions were
implemented for the angular coordinate $\theta$ in order to obtain an
eigenstate and orbital energy consistent with the variational
results~\cite{MocciaJCP40III}. This choice of boundary conditions,
that the derivative with respect to $\theta$ vanishes at the limits of
the mesh ($\theta = 0$ and $\theta = \pi$), leads to solutions
$\psi_{R(I)}(r,\theta)$ with a probability density consistent with the
$\theta$ dependence of the $3a_{1}$ orbital, as shown in
Figure~\ref{fig:density_contours}. The physical parameters of
interest, namely the resonance position, $E_{R}$, and width,
$\Gamma = -2E_{I}$, that characterize the tunneling process of the
quasi-stationary state when an external electric dc field is applied
along the $\pm\hat{z}$ directions, were found by solving
Eq.~(\ref{pde_system}) for a set of field strength values, $F_{0}$,
using a root search in order to find the energy that maximizes the
probability density amplitude in the $2d-$grid.

\section{\label{sec:results} Stark resonance parameters}

\begin{figure}[t!]
\includegraphics[width=0.49\textwidth]{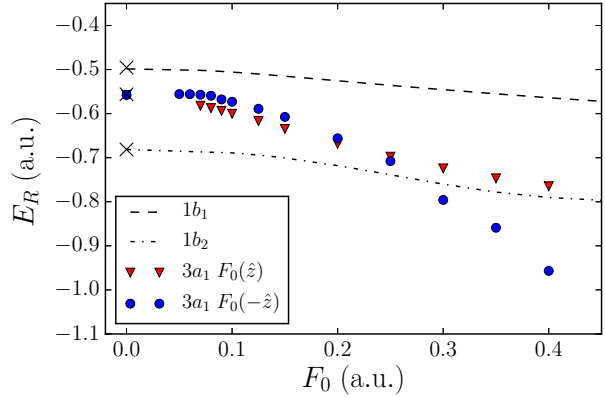}
\caption{Resonance position in atomic units as a function of the external
  field strength $F_{0}$ and the orientation of the field, along the
  $\pm\hat{z}$ direction (red triangles/blue circles), for the $3a_{1}$
  MO of H$_{2}$O. As a reference, the resonance position values for the $1b_{1}$
  (dashed line) and $1b_{2}$ (dot-dashed line) MO's are also included.}
\label{fig:pos3a1}
\end{figure}

\begin{figure}[t!]
\includegraphics[width=0.49\textwidth]{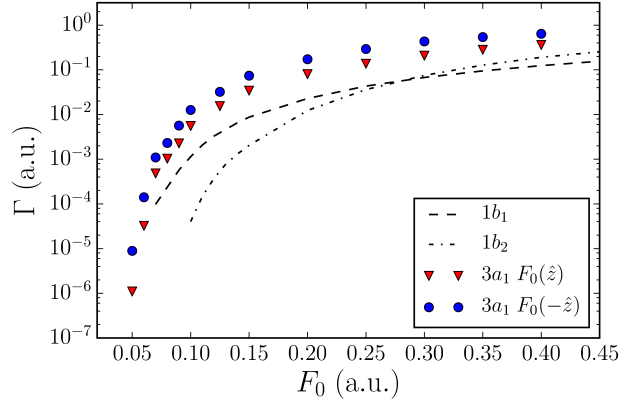}
\caption{Resonance width in atomic units as a function of the external
  field strength $F_{0}$ and the orientation of the field, along the
  $\pm\hat{z}$ direction (red triangles/blue circles), for the
  $3a_{1}$ MO of H$_{2}$O. For reference, the resonance widths for the
  $1b_{1}$ (dashed line) and $1b_{2}$ (dot-dashed line) MO's are also
  shown.}
\label{fig:width3a1}
\end{figure}

Results from applying the procedure described in
Section~\ref{sec:quantum_potential} are shown in
Figs.~\ref{fig:pos3a1} and~\ref{fig:width3a1}.

The resonance positions $E_{R}$ are shown in Figure~\ref{fig:pos3a1}
for external fields applied along the $\pm\hat{z}$ directions (red
triangles/blue circles) for a range of external field strengths. For
reference, the resonance positions obtained for the $1b_{1}$ and
$1b_{2}$ MO's using a spherically symmetric potential,
$V_{\rm{eff}}(r)$, are also indicated in the form of dashed and
dot-dashed lines respectively. For zero field strength $F_{0} = 0$
self-consistent eigenenergies obtained by Moccia~\cite{MocciaJCP40III}
are included as black crosses for the three valence orbitals of
interest. As expected, the resonance position for the $3a_{1}$ orbital
is bracketed by those for the $1b_{1}$ and $1b_{2}$ orbitals.

It can be noticed that for external fields applied along the
$-\hat{z}$ direction, where most of the density is located, the field
strength $F_{0}$ has to be strong, i.e., $F_{0}>0.1\ \rm{a.u.}$, for
the resonance position to change appreciably. On the other hand, the
resonance position for fields applied along $+\hat{z}$ appears to be
more sensitive at weaker fields. However the barrier appears to be
longer for external fields applied along the $+\hat{z}$ direction, at
a field strength of about $F_{0} = 0.25\ \rm{a.u.}$ the position
values cross, indicating a higher sensitivity of the resonance
positions for fields applied along the negative $\hat{z}$ direction as
the field strength is increased further.

Figure~\ref{fig:width3a1} shows the resonance widths corresponding to
external fields applied along the $\pm\hat{z}$ directions, as a
function of the field strength $F_{0}$. The results obtained with a
symmetric effective potential, $V_{\rm{eff}}(r)$, for the $1b_{1}$ and
$1b_{2}$ MO's are also shown as dashed and dot-dashed lines for
comparison purposes.

In analogy to the $m=\pm 1$ orbitals, the ionization rates for the
$3a_{1}$ MO, associated with the lifetime of the decaying state via
$\Gamma\tau=1$, exhibit a threshold behavior at the weaker field
strengths. Interestingly, for the two directions of the applied field,
we find a lower critical field strength for the $3a_{1}$ orbital in
comparison to what the more weakly bound orbital, $1b_{1}$, indicates.
In the tunneling region, the $3a_{1}$ orbital for fields applied along
the $-\hat{z}$ direction (blue squares) shows an ionization rate that
is about one order of magnitude larger than the ionization rate for
fields applied in the opposite direction (red triangles), this gap
becomes narrower as the field strength increases toward the
over-barrier regime.

\section{\label{sec:conc} Conclusion}

The Moccia single-center Hartree-Fock solution for the $3a_{1}$
orbital of H$_{2}$O has been investigated to understand its response
to a strong external dc electric field. We generalized a method to
obtain an effective potential to take into account $s-p$ type Slater
orbital mixing included in the Moccia orbital. We ignored small $l>2$
and particularly $m=2$ contributions to limit the form of the
effective potential to $V_{\rm{eff}}(r,\theta)$.

This permitted to study the relationship of the resonance parameters
(position and width) to the neighboring valence orbitals $1b_{1}$ and
$1b_{2}$ which were treated in a simplified approach before
($V_{\rm{eff}}(r)$ only, i.e., $1b_{1}\approx 2p_{x}$ and
$1b_{2}\approx 2p_{y}$). Interestingly, the $3a_{1}$ orbital is found
to ionize more easily than $1b_{1}$ or $1b_{2}$ irrespective of the
field direction along $\hat{z}$. The work should serve as motivation
for further studies of molecular orbitals of water using more
sophisticated wave functions.


\begin{acknowledgments}

  The financial support from NSERC of Canada is gratefully
  acknowledged.

\end{acknowledgments}

\bibliography{main}

\end{document}